\documentclass{article}
\usepackage[top=3cm, bottom=3cm, left = 3cm, right = 3cm]{geometry} 
\usepackage{amsmath,amssymb}  
\usepackage{bm}  
\usepackage{booktabs}
\usepackage{caption}
\usepackage{fancyhdr}
\usepackage{float}
\usepackage{graphicx} 
\usepackage[utf8]{inputenc}
\usepackage{memhfixc} 
\usepackage{multirow}
\usepackage{pdfsync}  
\usepackage{placeins}
\usepackage{textcomp}
\usepackage[pdftex,bookmarks,colorlinks,breaklinks]{hyperref}  

\newcommand{\keywords}[1]{\def\@keywords{#1}}

\begin{document}
\title{From Rapid Release to Reinforced Elite: Citation Inequality Is Stronger in Preprints than Journals}
\author{
 Chiaki Miura$^{1a*}$, Ichiro Sakata$^{a}$\\
    \small $^{1}$ORCID: 0009-0009-6492-0985 \\
    \small $^{a}$Department of Engineering/SkLab/Technology Management and Innovations,\\
    \small The University of Tokyo, Bunkyo, Tokyo 113-0033, Japan  \\
    \small $^{*}$Corresponding author: \tt{miura-tchiaki873@g.ecc.u-tokyo.ac.jp} \\
}
\date{This manuscript was compiled on \today}

\maketitle 


\begin{abstract}
 Preprints have been considered primarily as a supplement to journal-based systems for the rapid dissemination of relevant scientific knowledge and have historically been supported by studies indicating that preprints and published reports have comparable authorship, references, and quality. However, as preprints increasingly serve as an independent medium for scholarly communication rather than precursors to the version of record, it remains uncertain how preprint usage is shaping scientific discourse. Our research revealed that the preprint citations exhibit significantly higher inequality than journal citations, consistently among categories. This trend persisted even when controlling for age and the mean citation count of the journal matched to each of the preprint categories. We also found that the citation inequality in preprints is not solely driven by a few highly cited papers or those with no impact but rather reflects a broader systemic effect. Whether the preprint is subsequently published in a journal or not does not significantly affect the citation inequality. Further analyses of the structural factors show that preferential attachment does not significantly contribute to citation inequality in preprints, whereas author prestige plays a substantial role. Notably, the gap in citation inequality between the preprint category and the journal is more pronounced in fields where preprints are more established, such as mathematics, physics, and condensed matter physics.
\end{abstract}

\maketitle
\section*{Introduction}
The increasingly rapid transformations in modern society, coupled with the growing role of science, have elevated the importance of the rapid dissemination of scientific findings.
Preprints are intended to minimize the publishing delay due to article processing\cite{goldschmidt-clermontCommunicationPatternsHighEnergy2002} and have garnered significant attention during the COVID-19 pandemic, stimulating considerable debate\cite{kwonHowSwampedPreprint2020}.

While preprints have long been prominent in computer science, they are now playing a leading role across a broader range of disciplines.
As the post-publication peer review pipeline matures in processing and verifying more articles on preprint servers\cite{weissgerberAutomatedScreeningCOVID192021}, emergent publication platforms such as eLife and F1000, and publish-review-curate models\cite {eisenImplementingPublishThen2020} consider preprints as an independent, primary medium of academic discourse.
Although preprints contribute to accelerating science, researchers still debate whether they should treat them as equivalent to traditional publication channels.

Previous studies suggest that preprints and the corresponding journal articles are remarkably similar.
in authorship, reference\cite{AkbaritabarEtAl_StudyReferencingChanges_2022}, and qualitative expert evaluation.
However, even when such similarities exist, the way preprints are used may differ significantly from that of journals.
Many preprints are subsequently published in prestigious journals, such as Science and Nature Human Behavior, or in less rigorous and inclusive mega-journals, such as PLOS One and Scientific Reports.  
Articles that are published in prestigious journals have undergone multiple rounds of revision, are discussed among peers and at conferences, and are often presented as preprints as part of this process.
Therefore, the similarity in quality between preprints and their corresponding journal versions may be due to the fact that only articles of high quality are publicized as preprints, whilst those of lower quality are published in journals without the need for preprinting.
In order to demonstrate that the quality of preprints is equivalent to the corresponding version of record, we should measure how they impact a wide range of knowledge production, similar to how journal articles are now evaluated.

As preprints are considered of mixed quality, their use may be skewed towards a few reputable authors.
Inequality in citation distribution has already been observed in journals\cite{NielsenAndersen_GlobalCitationInequality_2021}, and this dynamic may be amplified in preprints. 
Unequal representation of articles hinders new theories and practices from gaining traction, impeding scientific progress\cite{chuSlowedCanonicalProgress2021}.
It is known that in an environment where actors lack prior knowledge about the validity of information, they disproportionately rely on reputable peers\cite{BendtsenEtAl_ExpertGameCooperation_2013}. 
The cultural diffusion model explains that conforming frequency-dependent copying significantly deforms the power-law distribution of trait frequencies\cite{mesoudiRandomCopyingFrequencydependent2009}.
Reference lists in one article are often directly transported to another\cite{macrobertsProblemsCitationAnalysis1989}, which may leave traces in citation distribution differently from other propagation of reference preference.
Citation network citing to and within the preprint system, mapped to the journal system via semantic similarity, can reveal the hidden selection bias in the system. This study aims to address this by examining differences in inequality and bias in citation between preprints and journals.

The explicit acknowledgment record of preprint use has 30 years of history and is short compared to the century-old practice of journal article reference\cite{_APAStyle_} and extensive analysis of knowledge diffusion\cite{priceNetworksScientificPapers1965}.
After an early experiment in the 1960s to share works via email\cite{Abelson_InformationExchangeGroups_1966}, preprint settled its way in the e-print and open access movement in the late 1990s\cite{Brown_EvolutionPreprintsScholarly_2001}, following the launch of the first widely recognized preprint repository arXiv started at Los Alamos National Laboratory in 1991.
The first attempt to quantitatively measure the scientific impact of preprints was conducted by\cite{LariviereEtAl_ArXivEprintsJournal_2014}.
They examined the citation to arXiv articles from papers indexed in Web of Science and found that the physics, mathematics, astronomy, and astrophysics arXiv versions are cited more promptly and decay faster than WoS papers. 
They reported that approximately one in fifteen references (6.6\%) in nuclear and particle physics is made to arXiv e-prints, while in the other specialties, 1.5\%.

\FloatBarrier

\section*{Result}\label{sec:result}
Citation inequality within each venue was quantified by the Gini coefficient $G$ calculated over five-year citation counts, $c_5$. 
\begin{align}
 G = \frac{1}{2 \bar{c_5}} \sum_{i=1}^{N} \sum_{j=1}^{N} |c_{5i} - c_{5j}|
\end{align}
where $\bar{c_5}$ denotes the mean $c_5$ of $N$ papers published in the venue.

\begin{figure}[t]
  \centering
  \includegraphics[width=1.0\linewidth]{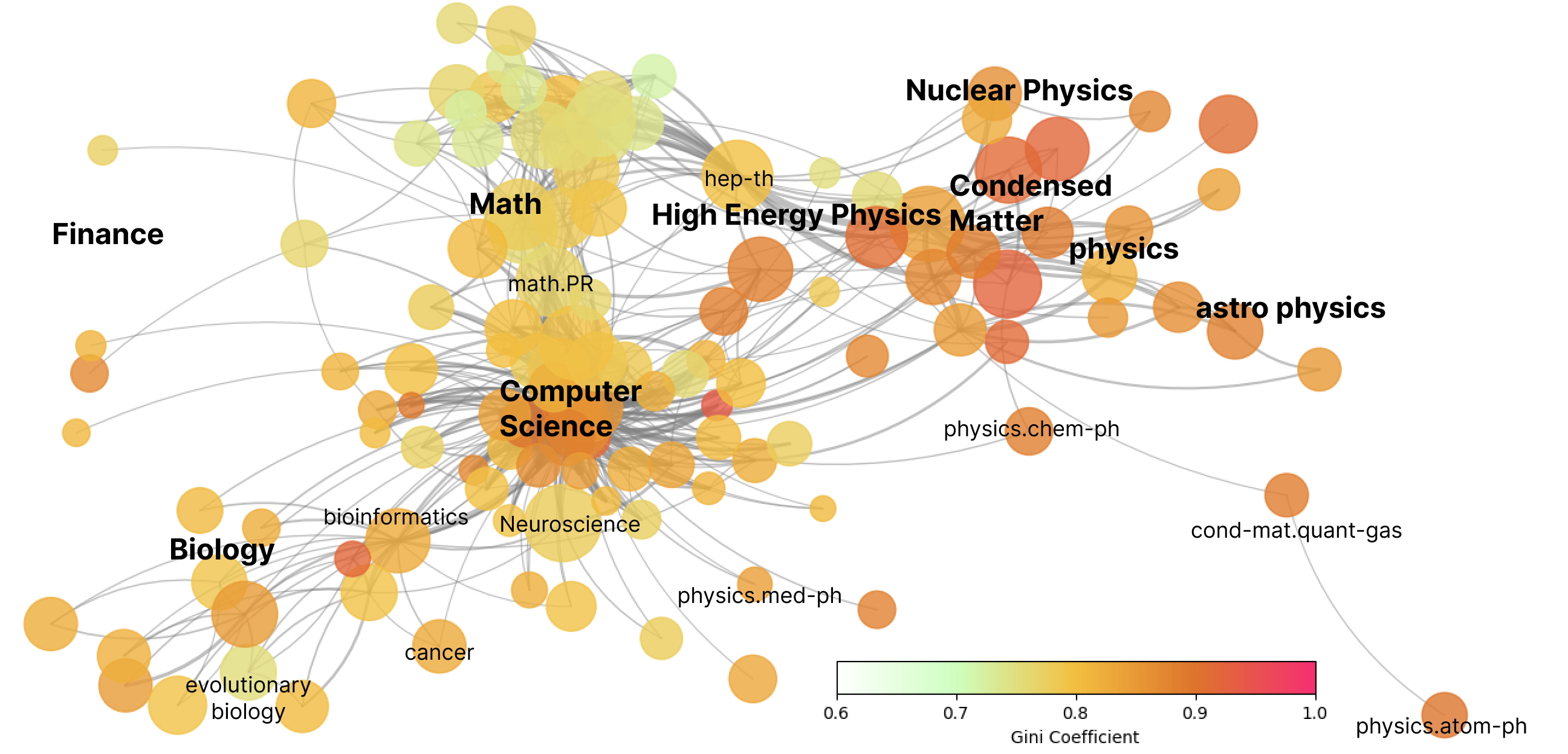}
  \caption{
 Citation inequality $G$ for each subcategory in arXiv and bioRxiv. The cool color indicates moderate inequality (blue$\approx 0.6$, green$\approx 0.7$), while warmer colors mean more substantial inequality (orange$\approx 0.9$, red means monopoly). 
 Subcategories are positioned more closely when more journal articles cite the same preprint in each category.
}
  \label{fig:citation_gini}
\end{figure}

Figure~\ref{fig:citation_gini} shows the strength of inequality for each preprint subcategory, which provides a natural, server-established resolution.
To illustrate how related subfields cluster, we positioned preprint categories in a co-journal citation network so that closely related categories appear near one another. 
Categories closer in the network are likely to exhibit similar citation inequality.

For each preprint category, we selected the ten most similar journals in terms of field, mean citation count $\bar{c_5}$, and the age $\tau$. 
Age is the number of years since the venue’s first publication.
Publication volume was excluded from the matching criteria, given that each bioRxiv category maintains a disproportionately larger output relative to the journal age.
Although population size remains a potential confounder, the variable of interest — the citation inequality — is theoretically invariant to the absolute size of the samples (see \ref{sec:matmeth} section 2). 
Based on the mean and variance of these journals’ Gini coefficients, we computed the normalized relative citation inequality $z$ for each preprint category.
\begin{align}
 z = \frac{G_{preprint} - \mu(G_{journal})}{\sigma(G_{journals})}
\end{align}
Here, $G_{preprint}$ is a Gini coefficient of a preprint category, $G_{journals}$ is a list of the coefficient for the matched journals, $\mu(\cdot)$ and $\sigma(\cdot)$ are the mean and standard deviation of a variable.

\begin{figure}[H]
  \centering
  \begin{minipage}{\textwidth}
    \centering
    \captionof{table}{Descriptive statistics of the preprint categories and their corresponding journals. We applied a logarithmic transformation to citation counts to normalize the distribution, reduce the impact of extreme values, and make the bin and matching more robust, which is denoted as $c^*$ in a table.}
    \label{tab:descriptive}
    \begin{tabular}{lrrrrrr}
      \multicolumn{1}{l}{} & \multicolumn{3}{c}{Preprint} & \multicolumn{3}{c}{Matched Journal} \\
      \cmidrule(lr){2-4} \cmidrule(lr){5-7}
 Category  & $\bar{c_5^{*}}$ & $\tau$& $N$  & $\bar{c_5^{*}}$ & $\tau$& $N$  \\
      \midrule
 astro-ph & 1.28 & 16 & 21622 & 1.32 (0.03) & 17.8 (1.1) & 462 \\
 behav & 1.14 & 11 & 25463 & 1.07 (0.12) & 13.7 (1.1) & 247 \\
 biomed & 1.10 & 11 & 11696 & 1.02 (0.11) & 13.6 (1.1) & 255 \\
 cond-mat & 1.10 & 29 & 31766 & 1.08 (0.06) & 29.8 (0.9) & 1650 \\
 cs & 1.51 & 27 & 116555 & 1.52 (0.04) & 28.6 (1.4) & 1348 \\
 econ & 1.32 & 8 & 330 & 1.25 (0.03) & 8.4 (1.3) & 64 \\
 eess & 1.33 & 8 & 1637 & 1.37 (0.03) & 7.9 (0.7) & 91 \\
 evol & 1.10 & 12 & 14880 & 1.05 (0.10) & 13.7 (1.1) & 237 \\
 hep & 1.32 & 34 & 37256 & 1.35 (0.03) & 33.0 (1.9) & 1158 \\
 math & 1.24 & 33 & 168088 & 1.23 (0.01) & 34.7 (3.5) & 627 \\
 molcel & 1.18 & 12 & 56732 & 1.05 (0.10) & 13.7 (1.1) & 237 \\
 nlin & 1.14 & 25 & 3797 & 1.04 (0.09) & 27.9 (1.1) & 623 \\
 nucl & 1.07 & 32 & 3530 & 1.07 (0.09) & 33.3 (2.1) & 357 \\
 physics & 1.12 & 29 & 26343 & 1.04 (0.08) & 31.4 (0.7) & 1226 \\
 q-bio & 1.17 & 22 & 9427 & 1.05 (0.10) & 23.8 (1.3) & 819 \\
 q-fin & 1.17 & 18 & 3331 & 1.15 (0.05) & 18.7 (0.9) & 210 \\
 stat & 1.56 & 18 & 17975 & 1.51 (0.04) & 19.7 (0.7) & 1007 \\
      \bottomrule
    \end{tabular}
  \end{minipage}
    
  \vspace{2em}

  \begin{minipage}{\textwidth}
    \centering
    \includegraphics[width=1.0\linewidth]{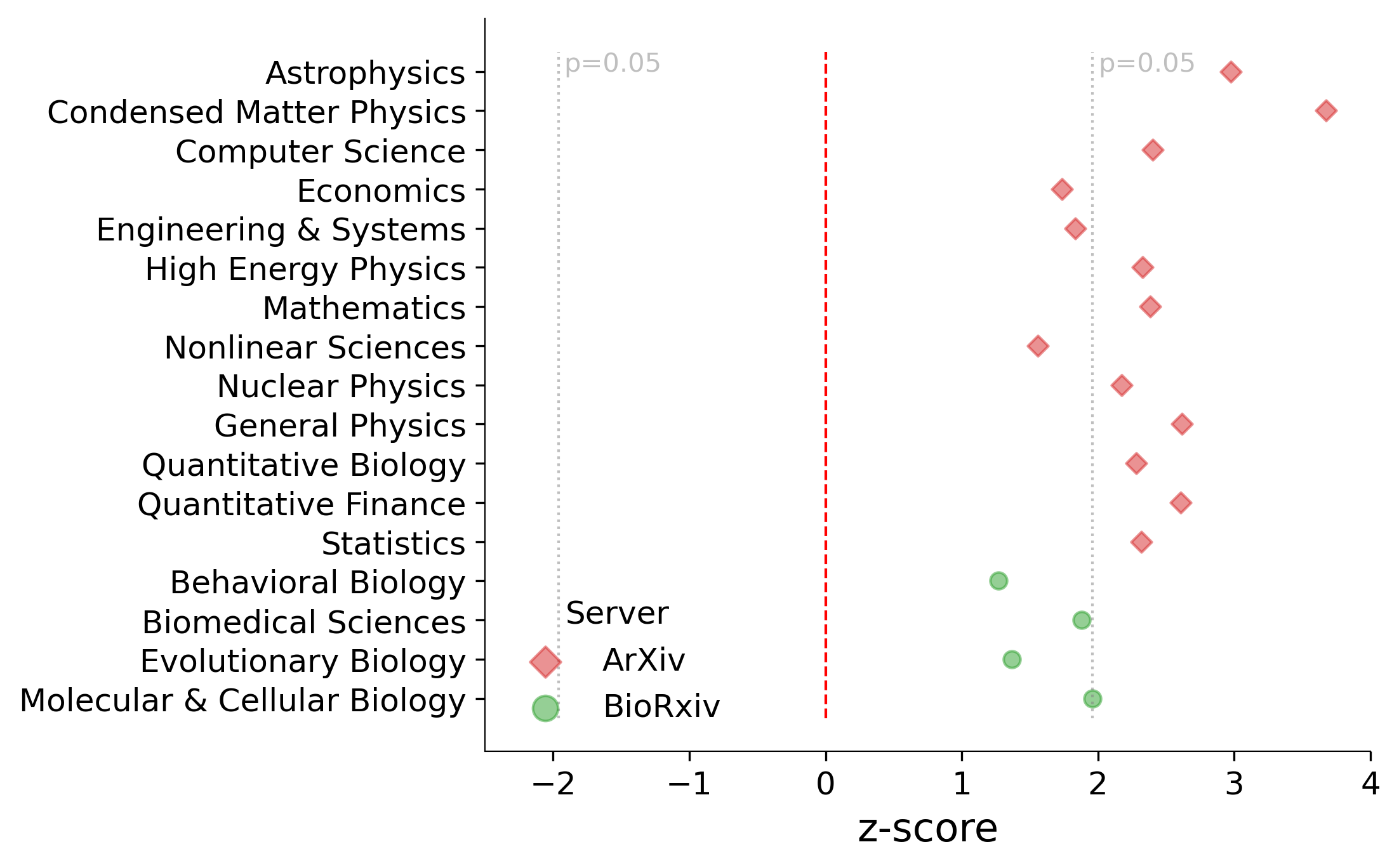}
    \caption{
 Each point represents a z-score compared to its control group, matched on journal age and mean citation. 
 A dashed line indicates no relative inequality.
 Dotted lines show 5\% significance threshold ($|z|$ = 1.96).
 }
    \label{fig:zscore}
  \end{minipage}
\end{figure}

The descriptive statistics for preprint coarsened exact matching to journals are shown in Table~\ref{tab:descriptive}.
In the following part, subcategories are aggregated into larger categories for concise representation.
Aggregation criteria are based on server-predefined categories (arXiv) or author-defined categories (bioRxiv), which result in 13 and 4 categories in arXiv and bioRxiv, respectively.

Across all categories, preprints exhibited $z>0$, which indicates higher levels of inequality (Figure~\ref{fig:zscore}).
We conducted the analysis both with and without trimming the top and bottom 1\% of values.
The trimmed analysis provided a conservative estimate, reducing the impact of potential outliers. In contrast, the full-data analysis captured the entire distribution, including the extremes that are an intrinsic part of the phenomenon under study. 
Both approaches point in the same direction, though the magnitude and statistical significance of the effect are more pronounced in the full data.

Surprisingly, despite a defined “cut-off” when preprints transition to journal publication—after which citations are expected to accrue to the journal version—citation inequality in preprints remains elevated. 
According to Xie et al.\cite{XieEtAl_PreprintFutureScience_2021}, the average time to journal publication is 406 days for bioRxiv and 224 days for arXiv.
By the time of our analysis, most preprints were older than five years and thus had already been curated.
Editors and reviewers redirect citations to the journal versions, shortening the window during which preprint citations can accumulate.
This limits the duration of preferential attachment dynamics, thereby reducing the extent to which preprints can accumulate disproportionate citation attention.


The highest relative citation inequality was observed in the field of condensed matter, followed by astrophysics, general physics, and quantitative finance, while the lowest relative inequality appeared in fields such as behavioral and evolutionary biology and non-linear sciences.
The Spearman’s rank correlation coefficient between the z-scores and raw Gini coefficients was approximately 0.38, indicating a moderate correlation.

To better understand these results, we examined two paths of reciprocal reputation reinforcement: preferential attachment and author journal prestige. 
Preferential attachment reflects how the citations are distributed based on existing popularity and thus intrinsic to the system’s citation dynamics, such as article search path and the process of gaining visibility, while the prestige of authors gained in traditional journals reflects social and reputational factors not encoded directly in the citation network, which is an external factor.

\FloatBarrier
\newpage

\subsection*{Preferential Attachment}
Preferential attachment refers to the tendency for an already popular network node to attract more links from other nodes.
Particularly in the journal article citation network, the number of citations gained in the next time unit is known to be linear to the number of citations already held\cite{JeongEtAl_MeasuringPreferentialAttachment_2003}.
Consider a probability $\Pi$ of a paper with a given citation count $c$ being cited in the next time unit.
The strength of preferential attachment is quantified by the exponent $\alpha$ in the following equation:
\begin{align}
  \Pi(c) = \frac{c^\alpha}{\sum_{i=1}^{N} c_i^\alpha}
\end{align}

\begin{figure}[t]
  \centering
  \includegraphics[width=1.0\linewidth]{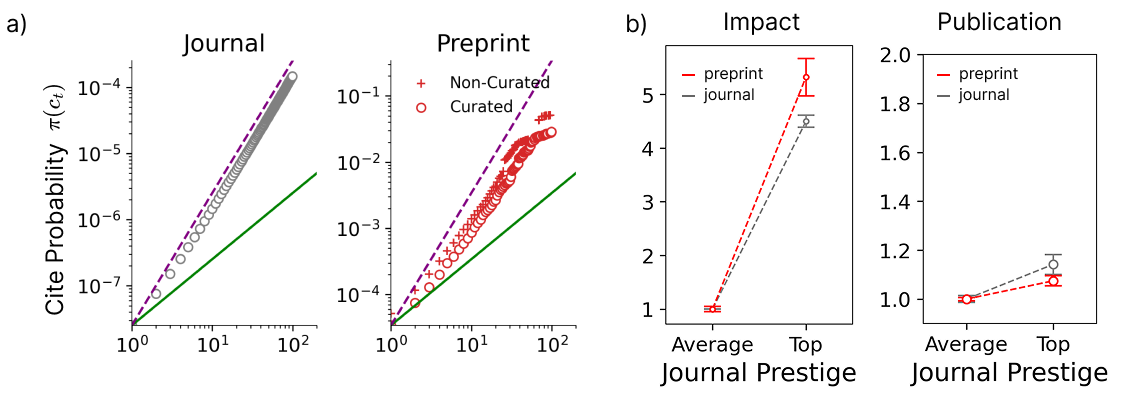}
  \caption{
 a) Preferential attachment for journals and preprints. The slope of the fitted line corresponds to the exponent $\alpha$.
 b) left: Author journal prestige and their relative impact on journals and preprints. Right: Author journal prestige and their relative publication on journals and preprints. In both panels, the error bar indicates the 95\% confidence interval.
 }
    \label{fig:pref_attch}
  \label{fig:auth_prstg}
\end{figure}

In Figure~\ref{fig:pref_attch}.a, we show the cumulative probability distribution $\pi$ for better visibility.
The slope of the fitted line of the plot corresponds to the exponent $\alpha$.
\begin{align}
  \pi(c) = \int_{0}^{c}\Pi(c) dc 
\end{align}
where $\alpha > 1$ indicates a superlinear relationship, which would yield a more highly skewed distribution than with lower $\alpha$.

We found that the preferential attachment in preprints was, at most, comparable to that in journals (Figure~\ref{fig:pref_attch}.a).
Curation of preprints, which is a publication in a journal, can affect the citation dynamics, but both curated and non-curated preprints exhibited similar preferential attachment.
If the observed citation inequality (raw $G_{preprint\_all} = 0.88$) were solely due to preferential attachment, the exponent $\alpha$ needs to be around 1.3, while the actual slope $\alpha<1.0$ (see \ref{sec:matmeth}).

\subsection*{Author Prestige}

Next, we compare the authors’ reputations in journals with their relative impact in preprints. 
We first took authors who have published in both preprints and journals.
Authors were divided into two groups—average and top— according to their impact from their journal publication.
After that, we compared them to the authors’ impact in preprint and journal publications between the strata.
For example, if the journal article of the top author is cited on average 100 times and preprinted 8 times, while that of the average author is cited 10 times and 1 time, respectively, the relative impact of the journal is 10, and that of the preprint is 8.

Suppose the preprint impacts are boosted by the author’s journal prestige. In that case, we must observe that the relative impact of preprint by the same top-tier authors is even higher compared to the relative impact gained from their journal publications.
In other words, if the preprint impact is affected by the author’s prestige, we expect that the citation of the top author’s preprint is boosted by some factor, widening the gap between the top and average authors.

We found that top journal authors have a significantly higher impact on preprints while producing less than average authors(Figure~\ref{fig:pref_attch}.b).
In other words, papers involving authors highly recognized in traditional journals are disproportionately likely to be cited than those not.
These findings indicate that researchers who are already influential in journals may exert even more substantial influence within preprint ecosystems.

Taken together, these trends suggest that preprints are not simply a rapid release of journal articles but are also more susceptible to the influence of elite researchers.

\section*{Discussion}\label{sec:discussion}
Preprints have long been proposed as a potential substitute for journals, and therefore, prior research focused on descriptive characteristics of preprint as a precursor in relation to journal articles.
However, given that a substantial number of preprints are expected to be eventually published in journals,
empirical similarities alone do not provide sufficient evidence that preprints and journals serve equivalent roles in the scholarly communication system.
To understand how the rise of preprints may transform practice, we need to shift our focus from the content of preprints to their usage patterns and how they form (and reform) science.
Our study finds that preprints exhibit universal systemic differences as a more substantial citation inequality in comparison to journal articles.
This pattern held even after acconting for the various confounding factors such as the age, the average citation count, visibility, and the scope of the journals/preprints.

Further analysis indicates that this reinforced inequality may be the result of social factors rather than intrinsic to the citation dynamics, such as indirect citation or reference copying.
Researchers cannot read all the papers and would rely on other articles they came across to choose which articles to read.
The process is the same for both preprints and journals, but when she cites a preprint, she would be more cautious about citing the preprint from a less-known author than she would when choosing journal articles.
Importantly, established researchers publishing in prestigious journals also gain higer visibility and citations, creating a similar kind of prestige feedback loop.
The interesting part here is that despite this commonality, the level of inequality is consistently higher in preprints than in journals across different fields of diffrent practice.

We might expect that in more preprint-established fields, the citation inequality would diminish to the same level as in journals.
However, Spearman’s rank correlation between the time $\tau$ since the preprint category’s first inception and the z-score is 0.59, indicating a moderate correlation, while that with the raw Gini coefficient is -0.06.
This suggests that the reinforced inequality is more prominent in the fields where preprints are more widely adopted.
In particular, fields such as computer science—where the preprint is deeply integrated—exhibit a higher concentration of impact even after excluding canonical outliers such as “Attention is All You Need.”
This may indicate that in these domains, the preprint ecosystem amplifies the visibility and advantage of already well-known authors, possibly due to the accelerated and broad dissemination mechanisms that preprints afford.

Field-specific analyses reveal that the relative citation imbalance is significant in physical sciences while less profound in the fields of behavioral and evolutionary biology, non-linear sciences, and economics.
Astrophysics and condensed matter science progress in a larger team and often rely on a shared dataset.
This may lead to a situation where a specific group of researchers is more likely to be cited.
Note that the analysis is limiting the maximum number of authors to 10, which mitigates the effect of kilo authors.

Another possible explanation is that the rigor of the journal peer review process plays a crucial role in the observed differences.
Prior studies show that the peer review length is shoreter in physical sciences and computer sciences than in economics.
Extended scrutiny can create a more equitable environment, as even renowned researchers must undergo the same waiting period for their results to be published.
In other words, it may not be that the imbalance in preprints is higher, but rather the inequality in journals is lower since many prominent researchers in the mature preprint ecosystem are more willing to publish their results on preprints if the preprint publishing is the best way to obtain broad exposure.
Two observations help clarify these dynamics. 
The number of journal articles citing preprints can be a proxy for the adoption of preprints and is comparable across fields. 
Meanwhile, although peer review duration may partly explain the observed differences, marked disparities are still seen in mathematics, which is known for its structured and rigorous review processes.

This study has several limitations. 
For one, the sheer volume of preprint publications did not allow us to match any preprint categories with journals, which may introduce an indirect impact on citation inequality.
Publication rate affects the amount of attention each article receives, and given that the number of articles authors read cannot be ten times more than it was in the three decades before. Naturally, the ratio of articles receiving any attention will be lower.
Therefore, outcomes may partially reflect underlying differences in the publication rate.
For two, our analysis ignored the impact through other outlets, such as conferences and social media. 
Preprints are often publicized right before or after they are presented at conferences, and ideas exchanged and discussed there stimulate the work that eventually leads to citing the work.
Here, the acceptance rate is the confounder.
Prominent researchers have a higher chance of acceptance and can enhance their visibility, and the effect diminishes at the time of journal publication.

Preprints remain an essential infrastructure for accelerating science by allowing researchers to disseminate findings, eliminating processing delays.
Our results present several policy implications for the future of preprints.
First, as preprints increasingly fulfill roles traditionally held by journals, the tendency for prestige to concentrate on a few individuals can demolish the research diversity.
Uncovering the underlying structure of imbalance is needed to calibrate evaluation metrics accordingly when incorporating preprints as a measure of research excellence.
Second, the peer review process is not only a quality control mechanism to validate articles but a social mechanism to distribute attention to an article to less known authors through the thorough scrutiny that every article should pass.
If the initial visibility imbalance results in citation inequality, curation of the preprint is not the solution.
Given the empirical evidence that disruptive works often remain unnoticed\cite{GolosovskyLariviere_UncitedPapersAre_2021, miuraLargescaleAnalysisDelayed2021}, another way to redistribute the visibility is needed.
Finally and most importantly, analyzing preprint from the use perspective can provide rich information on how rapid dissemination accelerates scientific progress.
Nevertheless, the research is few and far between on how preprints integrated into scientific knowledge production as a distinct pipeline.
Emergent social issues pose a trade-off between efficient knowledge flow and sufficient scrutiny.
It is crucial to distinguish whether the observed asymmetry in citation patterns stems from the structural nature of science—where a few highly cited breakthroughs rapidly emerge from a broad base of inconclusive or low-impact studies—or merely from the amplification of reputation in unvetted environments. 

\section*{Acknowledgements}
This work was supported by JST, ACT-X Grant Number JPMJAX24CP, Japan.
This work will be presented at the 20th INTERNATIONAL CONFERENCE ON SCIENTOMETRICS \& INFORMETRICS (ISSI 2025).

\bibliography{bib.bib}

\newpage

\section*{Materials and Methods}\label{sec:matmeth}
\subsection*{Data collection, normalization, and matching}
We combined the world’s largest bibliographic database, OpenAlex, with the snapshot of the two largest preprint servers, arXiv and bioRxiv.
Both preprint servers have distinct publication practices, with arXiv being a preprint server for physics, mathematics, and computer science, taken top three major categories by publication, while bioRxiv is focused on biology, which together covers various fields of study in physical, formal, biological sciences and part of life sciences.
We first obtained the preprint snapshot for arXiv from Geiger (2019)\cite{Geiger_ArXiVArchive_2019} and for bioRxiv via the official API.
Gieger (2019) contains a total of 1,635,684 category-wise and year-wise metadata ranging from 1993 to 2018, from which we extracted the articles published between 2015 and 2019, amounting to 637,944 articles.
BioRxiv API provides metadata for all the 262,899 articles published, and we fetched those published between 2015 and 2019, which amount to 261,904 articles.
Articles from both sources were reduced to the first version of the individual report.
We then match them to OpenAlex records by doi, primary location source id (“S4306400194” and “S4306402567”), and raw PDF URL string, yielding 426,448(66.8\%) and 67,274(25.7\%) preprints on arXiv and bioRxiv, respectively.
OpenAlex metadata is used to obtain the reference and citing articles of the preprints, disambiguate authors, and match preprint categories to journal subfields.
We used OpenAlex Snapshot, taken on 1 May 2025, as the primary dataset, which sufficiently contains the articles published until the previous year.

We restricted the analysis to articles and excluded other types such as conference proceedings, book chapters, reviews, erratum, and editorials to mitigate the possible biases.
Also, post-prints were excluded from the analysis as they can have different publication motivations and citation practices\cite{_NIHPublicAccess_}.
All the analyses consider journal and preprint articles published between 2015 and 2019 unless stated otherwise.
This is to mitigate the effect of citation inflation\cite{PetersenEtAl_MethodsAccountCitation_2019} and other year-fixed effects, including the effect of the preprint publication surge in the COVID-19 pandemic.

Next, we matched preprint categories to journals of the shared research scope and compared the citation inequality between preprints and journals. 
The matching criteria for journals were the scope of the venue, the mean citation count $\bar{c_5}$, and the age $\tau$ of the journal.
The scope of the venue is defined by OpenAlex’s primary topic subfield assigned to each article.
Aggregating the number of articles in each subfield, we selected the three most frequent subfields for each venue, which are the most representative of the venue.
A journal is considered to be in the same scope if it shares at least one of the three representative subfields of the preprint category.
The mean citation count is calculated after winsorizing the citation count at 1\% on both ends of the distribution to reduce the influence of extreme values.
Also, when matching journals, we took the log of the mean citation count to address the issue of high variability within the variables.
This transformation helps to normalize the distribution and make relationships clearer. 
Journal ages $\tau$ are inferred from the first year $y$ with a noticeable publication threshold $N_0$, where we took $N_0=10$ for our analysis.
The publication threshold is to eliminate the possible bias of database misassignment of the publication date.
We calculated the age as $\tau = y_0 - y$, where the base year is set to 2025 ($y_0=2025$).

\subsection*{Preprint categories}
ArXiv categories are defined by two levels, separated by a dot, such as astro-ph.CO, where we define major categories as the first part (astro-ph) and subcategories as the second part (CO).
The exception is made for the following categories: acc-phys, adap-org, alg-geom, ao-sci, bayes-an, chao-dyn, cmp-lg, comp-gas, dg-ga, funct-an, mtrl-th, patt-sol, q-alg, and solv-int are ignored as they are the predecessor of another category and no articles fall within the specified time frame.
BioRxiv subcategories are those defined by the preprint servers, and major categories are those aggregated by the authors.
The aggregation table is shown in Table~\ref{tab:aggregation}.

\begin{table}[h]
  \centering
  \caption{Aggregation of preprint categories.}
  \label{tab:aggregation}
  \begin{tabular}{ll}
    \toprule
 Preprint Category & Aggregated Category \\
    \midrule
 animal-behavior-and-cognition & \multirow{4}{*}{behavioral biology} \\
 developmental-biology & \\
 neuroscience & \\
 zoology & \\
    \midrule

 cancer-biology & \multirow{7}{*}{bio-medical science} \\
 clinical-trials & \\
 epidemiology & \\
 immunology & \\
 pathology & \\
 pharmacology-and-toxicology & \\
 physiology & \\
    \midrule
    
 ecology & \multirow{4}{*}{evolutionary biology} \\
 evolutionary-biology & \\
 paleontology & \\
 plant-biology & \\
    \midrule

 bioengineering & \multirow{12}{*}{molcular and cellular biology} \\
 bioinformatics & \\
 systems-biology & \\
 biophysics & \\
 synthetic-biology & \\
 biochemistry & \\
 cell-biology & \\
 genetics & \\
 genomics & \\
 molecular-biology & \\
 microbiology & \\
    \midrule
  \end{tabular}
\end{table}

\subsection*{Citation inequality metrics and its sample size independence}
Citation inequality within each venue was quantified by the Gini coefficient $G$ calculated over five-year citation counts, $c_5$. 
We chose the Gini coefficient for its scale-independence and normalization properties, which allows for comparison across different fields.
The metric should be interpreted carefully as it is sensitive to extreme values, and identical values can arise from different distributions.
To mitigate the marginal bias, we excluded articles with extreme citation counts, 1\% at both ends of the distribution.
Citation counts are known to follow a lognormal distribution in both journals\cite{wangQuantifyingLongTermScientific2013} and preprints\cite{FraserEtAl_RelationshipBioRxivPreprints_2020}, indicating a shared underlying form. 

Citation inequality (Gini coefficient) $G$ is an area between the perfect equality line and lorentz curve $L(\cdot)$.

\begin{align}
 G = \int_{0}^{1} \Bigl( x - L(x) \Bigr) dx = 1 - 2 \int_{0}^{1} L(x) dx
\end{align}

The Lorenz curve $L(p)$ at probability $p\in(0,1)$ is, by definition,
\begin{align}
 L(p)&\equiv \frac{1}{\mathbb{E}[X]}
       \int_{0}^{F_X^{-1}(p)} x f_X(x) dx \quad, 
\end{align}
where $F_X^{-1}(p)$ is the $p$-quantile of $X$.

Given that the citation is distributed lognormally within a journal, we can write a citation distribution as:
\begin{align}
 f_X(x)  &=\frac{1}{x\,\sigma\sqrt{2\pi}}\exp\Bigl(-\frac{(\ln(x)-\mu)^2}{2\sigma^2}\Bigr)
\end{align}

Letting $Y=\ln(X)$, we can rewrite the integral part.

\begin{align}
     \int_{0}^{F_X^{-1}(p)} x\,f_X(x)\,\mathrm{d}x
     &= \int_{-\infty}^{y_p}
 e^y\,f_Y(y)\,\mathrm{d}y, \\
    &= \exp\Bigl(\mu+\tfrac12\sigma^2\Bigr)
     \,\Phi\!\Bigl(\tfrac{y_p-(\mu+\sigma^2)}{\sigma}\Bigr).
   \end{align}
 where $y_p=\ln\bigl(F_X^{-1}(p)\bigr)$, $\Phi(\cdot)$ is the standard normal CDF, and $\Phi^{-1}(\cdot)$ is its quantile. 

Dividing the integral term by $\mathbb{E}[X] = \exp(\mu+\tfrac12\sigma^2)$ cancels out that exponential factor. Since $p=F_X\bigl(e^{y_p}\bigr)=\Phi\!\bigl(\tfrac{y_p - \mu}{\sigma}\bigr)$, we have  $\,y_p=\mu + \sigma\,\Phi^{-1}(p).$ Substituting $y_p$ back gives
\begin{align}
 L(p)\;=\;\Phi\!\Bigl(\,\Phi^{-1}(p)\;-\;\sigma\Bigr)
\end{align}

\begin{align}
G = 1 - 2 \int_{0}^{1} L(p) dp = \Phi\!\Bigl(-\,\tfrac{\sigma}{\sqrt{2}}\Bigr)
\end{align}
\begin{align}
 \boxed{ G = \Phi\!\Bigl(\tfrac{\sigma}{\sqrt{2}}\Bigr) }
\end{align}

That is, for a lognormal random variable $X$ with parameters $(\mu,\sigma)$, the Lorenz curve depends only on $\sigma$, and thus $G$ is scale-agnostic and sample-size independent.\\

\subsection*{Preferential attachment}
We estimated the preferential attachment exponent $\alpha$ by fitting a linear model to the log-log plot of cumulative citation probability $\pi(c)$ against $c$. 
Citation counts were taken at 4 years from publication and the following year, matching the citation window used in the Gini coefficient analysis.

\begin{figure}[t]
  \centering
  \includegraphics[width=0.8\linewidth]{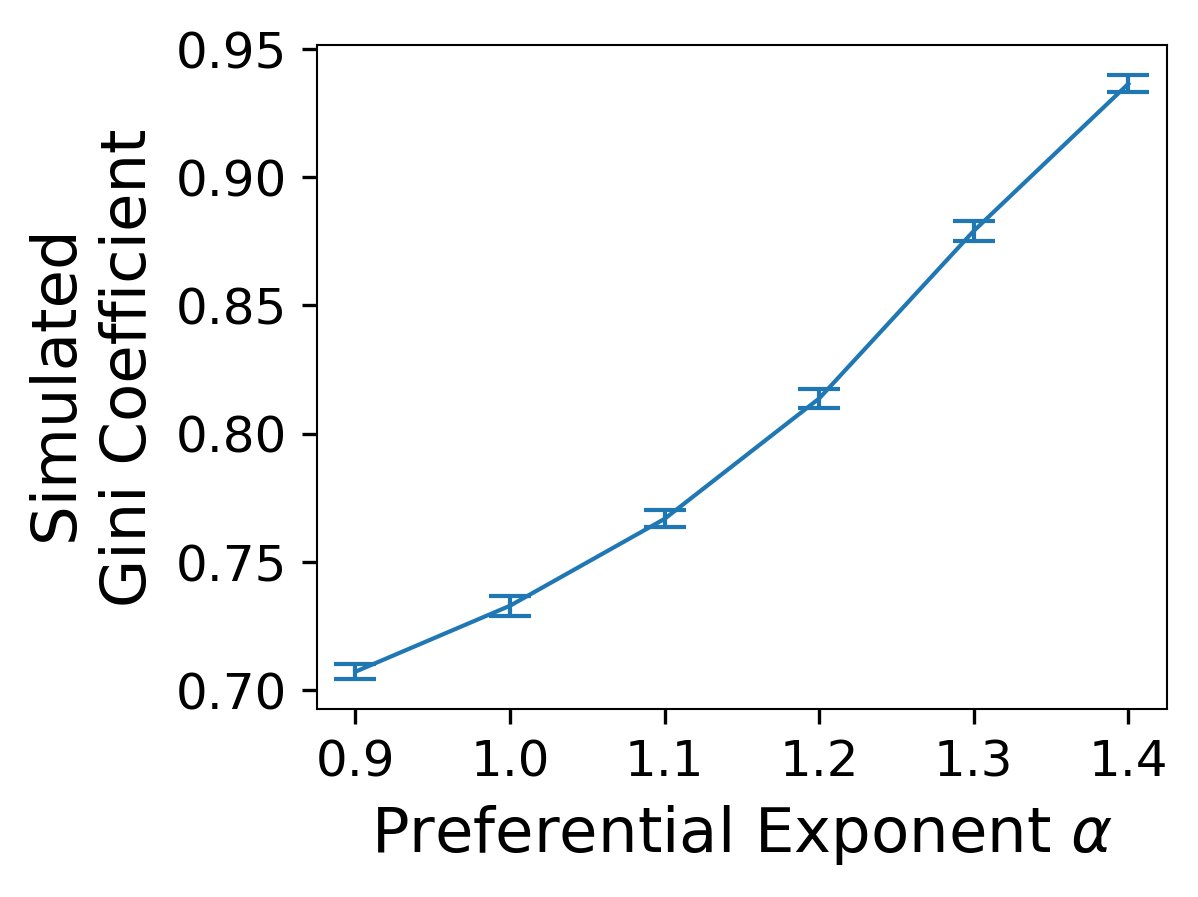}
  \caption{The simulation of preferential attachment. The slope of the fitted line corresponds to the exponent $\alpha$.}
  \label{fig:pref_sim}
\end{figure}
From numerical simulation on a finite sample of paper, we can estimate the lower threshold of the exponent $\alpha$ needed to exhibit a certain level of inequality $G$ (Figure~\ref{fig:pref_sim}.)
The simulation is run on a Barabasi-Albert model\cite{JeongEtAl_MeasuringPreferentialAttachment_2003} on $m$ nodes clique graph with $m=4$; each runs $5\times10^3$ iterations with a different random seed.
The simulation is repeated 30 times, and the mean and standard deviation of the exponent $\alpha$ are calculated.

\subsection*{Author journal prestige}
We selected 935,363 authors who had published at least one preprint and one journal article.
These authors were divided into two groups based on their journal impact.
Top authors were defined as those whose journal articles ranked within the top 10\% by mean Field-Weighted Citation Impact (FWCI)\cite{WaltmanvanEck_SystematicEmpiricalComparison_2013}.
FWCI scores were obtained from OpenAlex for both preprint and journal articles published between 2015 and 2019.
To compare the relative publication volume between the top and average authors, we computed the total number of preprints and journal articles published by each author group in the analysis period (2015-2019).

The relative impact of the journal is calculated as the ratio of the mean FWCI of the top authors to that of the average authors.
After the mean FWCI and the 95\% confidence interval of the top authors are calculated, we divide them by the mean FWCI of the average authors.
Similarly, the relative impact of preprint is calculated as the ratio of the mean FWCI to that of the average authors.

\end{document}